\documentclass[10pt,twocolumn,letterpaper]{article}

\usepackage[pagenumbers]{iccv} 
\usepackage{pifont}
\definecolor{iccvblue}{rgb}{0.21,0.49,0.74}
\usepackage{amsmath, bm}
\usepackage{mathtools}
\usepackage{comment}
\usepackage{multirow}
\definecolor{brightred}{RGB}{230, 0, 0} 
\definecolor{brightblue}{RGB}{0, 0, 230} 
\usepackage[pagebackref,breaklinks,colorlinks]{hyperref}
\hypersetup{
    linkcolor=Red,           
    citecolor=iccvblue,         
    filecolor=Magenta,       
    pagebackref=red,
}

\def\paperID{6366} 
\def\confName{ICCV}
\def\confYear{2025}

\title{FB-Diff: Fourier Basis-guided Diffusion for Temporal Interpolation of 4D Medical Imaging}

\author{
  Xin You$^{1,2,3}$ \quad Runze Yang$^{4}$ \quad Chuyan Zhang$^{2}$ \quad Zhongliang Jiang$^{1}$ \quad Jie Yang$^{2}$ \quad Nassir Navab$^{1}$ \\
  \\
  $^1$ Computer Aided Medical Procedures,
Technical University of Munich, Munich, Germany \\
  $^2$ Institute of Medical Robotics, Shanghai Jiao Tong University, Shanghai, China \\
  $^3$ Munich Center for Machine Learning, Munich, Germany \\
  $^4$ School of Computing, Macquarie University, Sydney, Australia \\
  {\tt\small xin.you@tum.de, nassir.navab@tum.de}
}

\begin{document}
\maketitle

\begin{abstract}
\noindent The temporal interpolation task for 4D medical imaging, plays a crucial role in clinical practice of respiratory motion modeling. Following the simplified linear-motion hypothesis, existing approaches adopt optical flow-based models to interpolate intermediate frames. However, realistic respiratory motions should be nonlinear and quasi-periodic with specific frequencies. Intuited by this property, we resolve the temporal interpolation task from the frequency perspective, and propose a \textbf{F}ourier \textbf{b}asis-guided \textbf{Diff}usion model, termed FB-Diff. Specifically, due to the regular motion discipline of respiration, physiological motion priors are introduced to describe general characteristics of temporal data distributions. Then a Fourier motion operator is elaborately devised to extract Fourier bases by incorporating physiological motion priors and case-specific spectral information in the feature space of Variational Autoencoder. Well-learned Fourier bases can better simulate respiratory motions with motion patterns of specific frequencies. Conditioned on starting and ending frames, the diffusion model further leverages well-learned Fourier bases via the basis interaction operator, which promotes the temporal interpolation task in a generative manner. Extensive results demonstrate that FB-Diff achieves state-of-the-art (SOTA) perceptual performance with better temporal consistency while maintaining promising reconstruction metrics. Codes are available at \url{https://github.com/AlexYouXin/FB-Diff}

\end{abstract}    
\section{Introduction}
\label{sec:intro}

\begin{figure*}[!t]
\centerline{\includegraphics[width=0.93\linewidth]{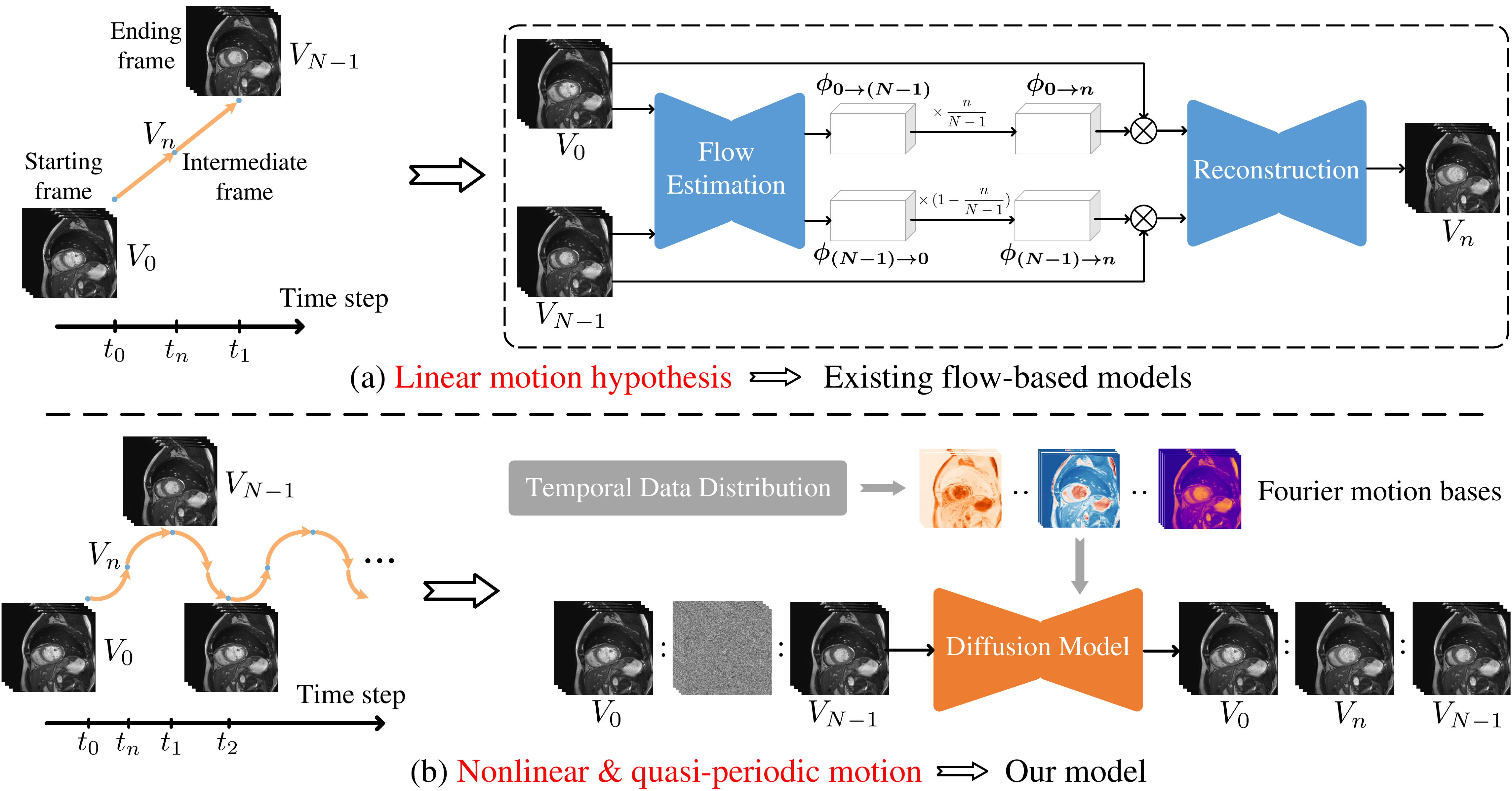}}
\caption{Temporal Interpolation for respiratory motions. (a) Simplified linear-motion hypothesis: interpolating intermediate frames by regressing optical flows with linear variations. (b) Nonlinear \& quasi-periodic motions: leveraging diffusion models conditioned on Fourier motion bases, which represent various motion patterns indicating nonlinear and locally inconsistent anatomical variations.}
\label{comparison}
\end{figure*}

4D medical images, which depict 3D volumes with temporal variations, are essential in clinical practice for capturing dynamic changes and monitoring the progression of diseases over time \cite{ehrhardt20134d, jeung2012myocardial, wang2009dosimetric, you2023semantic}. A significant application is the temporal modeling for breathing-induced motions of cardiac or pulmonary anatomical structures. That task aims to enrich motion visualizations by generating intermediate frames under the condition of starting and ending frames, which is consistent with the concept of Video Frame Interpolation (VFI) in natural scenarios.

VFI has been thoroughly investigated through systematic studies in recent years \cite{meyer2018phasenet, niklaus2020softmax, park2023biformer, zhang2023extracting, zhou2023exploring, kong2022ifrnet, danier2024ldmvfi, zhong2024clearer, jain2024video}. However, due to unique properties and constraints of medical imaging, it remains a challenge to directly implement these frameworks into 4D medical video interpolation with maintaining equally promising performance. Specifically, the quality of ground truth intermediate frames in medical imaging is often compromised due to factors such as imaging noise, unstable breathing, causing unexpected artifacts \cite{jeung2012myocardial, hor2011magnetic, smiseth2016myocardial, wang2009dosimetric}. Thus, the lower image quality compared to videos in the  natural domain, will degrade the interpolation performance of VFI models. Besides, potential patients' movement might induce an additional disturbance to the temporal modeling \cite{mizuno2021preoperative, caines20224dct}.


Furthermore, there exists a discrepancy on the type of motions between natural and medical scenes. The former focuses on the object movement and illumination intensity change. While in the medical domain, motions can be depicted as subtle anatomical variance with quasi-periodic motion discipline \cite{wei2023mpvf}. Thus, a specific motion prior of anatomies is effective to boost the temporal interpolation, especially for the medical data with limited resources from a small pool of individuals. 

On account of these challenges, we present the following question: ``How can we devise a specialist VFI model tailored for 4D temporal medical volumes?'' Existing methods \cite{guo2020spatiotemporal, wei2023mpvf, kim2024data, kim2022diffusion, balakrishnan2019voxelmorph} on medical VFI face a major shortcoming: the linear motion hypothesis in a breathing period is required for the regression of intermediate frames as revealed in Fig.\ref{comparison}(a). However, respiratory motions should be a nonlinear oscillatory process that cannot be simply modeled with a linear equation \cite{wei2023mpvf, li2024cpt}. Also, these methods fail to recognize that respiratory motions in heart or lung, are indeed a quasi-periodic motion process with specific frequencies by extending the timeline. Intuited by this, we resolve the respiratory motion simulation from a frequency perspective. In Fig.\ref{comparison}(b), each medical volume in the temporal domain can be viewed as a specific value in the periodic curve. That periodic curve can be decomposed as a group of sine and cosine bases with multiple frequencies \cite{tolstov2012fourier}. Comparatively, there exists a group of Fourier bases, which are capable of reconstructing original temporal sequences. Here Fourier bases with different frequencies refer to various motion patterns, which indicate nonlinear and locally anisotropic variations of anatomical structures. In our work, Fourier bases are acquired by the Fast Fourier Transform (FFT) on the temporal domain.

In this work, 4D data distributions of the respiratory process depict a regular motion discipline, with temporal variations from inhaling to exhaling, or from diastole to systole. Motivated by that observation, physiology motion priors are introduced to encode generic characteristics of the respiratory process. Then a Fourier Motion Operator $\mathcal{O}_{FM}$ is elaborately devised to yield enhanced motion bases by incorporating physiology motion priors and case-specific spectral information in the feature space. Specifically, A Variational Autoencoder (VAE) is introduced to map the video space into the feature space. Physiology motion priors are defined as learnable frequency embeddings, which can be updated by pretraining a Variational Autoencoder (VAE) on given temporal data. And case-specific frequency information can be extracted by FFT in the temporal domain. Well-learned physiology motion priors can enrich Fourier bases with generic properties of respiratory motions. Then enhanced Fourier motion bases can better simulate the quasi-periodic respiratory process with motion patterns of specific frequencies. On the grounds that diffusion models show promising performance in generating realistic videos \cite{ho2020denoising, ho2022video, liu2024sora}, a Markov chain process is learned to transform Gaussian noise into target videos by the proposed \textbf{F}ourier  \textbf{B}ases-guided \textbf{Diff}usion, termed \textbf{FB-Diff}. Since only starting and ending frames are available during inference, they serve as conditional prompts for intermediate frame interpolation. Meanwhile, we devise the basis interaction operator to inject enhanced Fourier bases into the conditional diffusion model. Then, FB-Diff can well address the temporal interpolation task in a generative manner.

\begin{figure*}[!t]
\centerline{\includegraphics[width=0.93\linewidth]{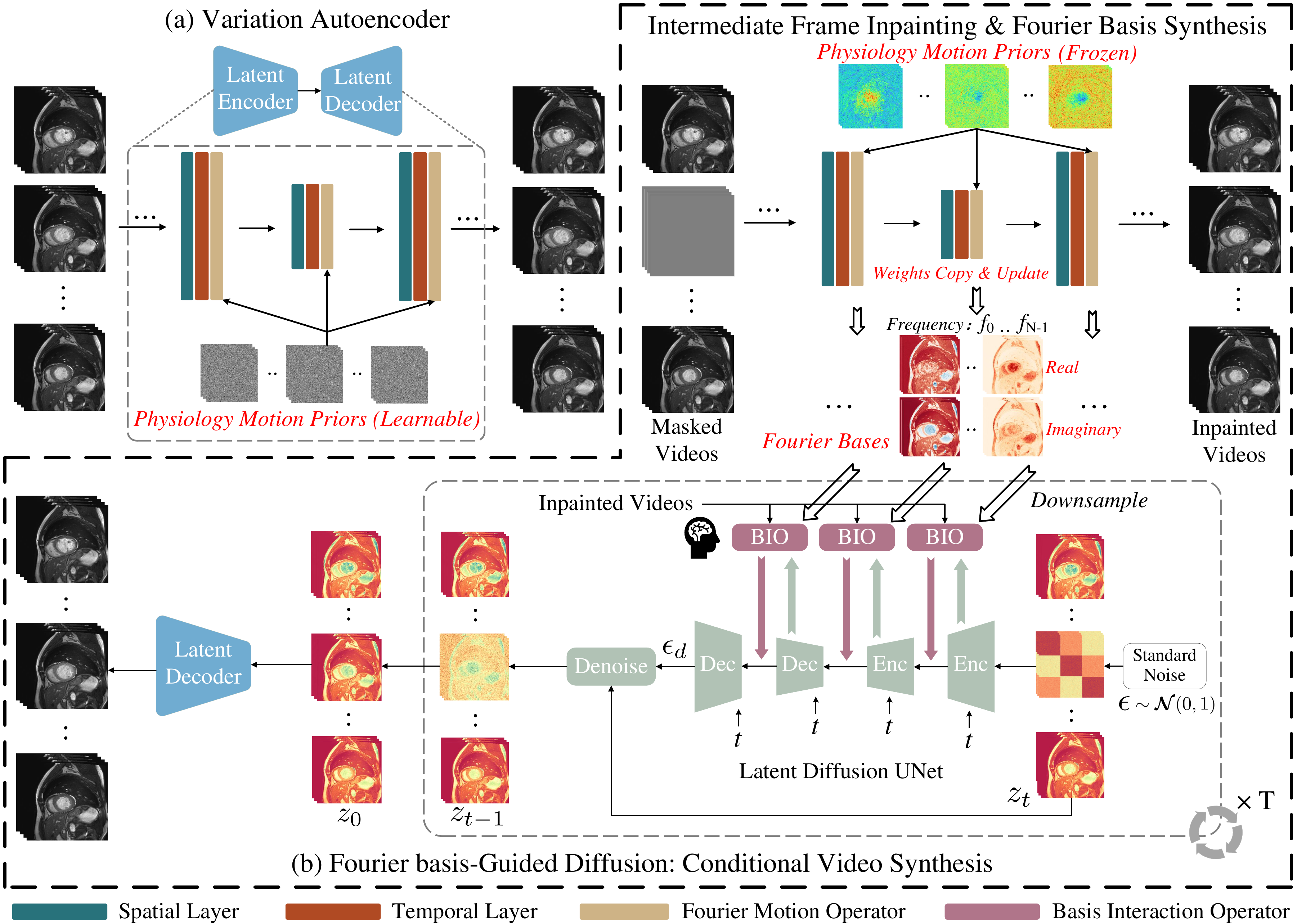}}
\caption{The pipeline of FB-Diff. (a) The Fourier motion operator is proposed to extract Fourier bases in a feature space of VAE. Bases consist of learnable generic motion priors and case-specific frequency information. (b) Fourier bases are first adapted to the masked video domain under the guidance of well-learned motion priors and pretrained VAE weights. Domain-adapted bases serve as the conditional input to boost the video synthesis by diffusion models.}
\label{pipeline}
\end{figure*}

Our proposed method has achieved state-of-the-art (SOTA) interpolation performance compared with other methods for 4D medical imaging, particularly on the midmost frames. For quantitative results, more comprehensive evaluation metrics are adopted in this work, including traditional reconstruction metrics and feature-level perceptual metrics. FB-Diff not only presents promising reconstruction metrics but also superior perceptual results. Qualitative visualizations reveal better temporal consistency by FB-Diff compared with other linear-motion-based methods. Our main contributions are summarized as follows:

\begin{itemize}
\item We propose a diffusion model termed FB-Diff for the temporal interpolation of 4D medical imaging. FB-Diff leverages Fourier motion bases, which indicate motion patterns with specific frequencies. Those bases can help to resolve the respiratory modeling with nonlinear and quasi-periodic motions.
\item We introduce the Fourier motion operator $\mathcal{O}_{FM}$ to extract Fourier bases, containing physiology motion priors and case-specific frequency representations.
\item We devise the basis interaction operator to inject Fourier bases into the diffusion model. Fourier bases, serving as the conditional guidance, can boost the video synthesis for temporal interpolation.
\item Extensive results demonstrate that FB-Diff reveals SOTA perceptual performance with better temporal consistency while maintaining promising reconstruction metrics.
\end{itemize}

\section{Related Work}
\textbf{Video Frame Interpolation.}
Popular VFI techniques are mainly classified into motion-free and motion-based approaches, depending on whether optical flows are incorporated or not. \textbf{Motion-based:} These approaches warp input frames forward or backward based on optical flows acquired by either off-the-shelf networks \cite{dosovitskiy2015flownet, ilg2017flownet, sun2018pwc, teed2020raft, wang2024sea} or custom-designed flow estimators \cite{zhang2023extracting, huang2022real, li2023amt, kong2022ifrnet, jin2023unified, hu2024iq}. Warped intermediate frames are then refined by delicately designed networks \cite{bao2019memc, jiang2018super} to enhance visual quality. Recently, Transformer-based motion estimators are proposed to raise the upper limit of VFI performance \cite{lu2022video, park2023biformer, zhang2023extracting} \textbf{Motion-Free:} Motion-free methods rely on
implicit spatio-temporal modeling \cite{choi2020channel,choi2021motion,hu2022spatial,kalluri2023flavr} to generate the intermediate frame. However, due to a limited quality of 4D medical videos and quasi-periodic motion priors, existing VFI frameworks are not feasible. That is why we aim to propose a specialist VFI model for 4D medical sequences.


\vspace{1mm}
\noindent \textbf{4D Medical VFI.} 
Existing methods on this topic mainly leverage volumetric motions for 4D temporal interpolation. Specifically, SVIN \cite{guo2020spatiotemporal} estimates forward and backward deformation fields, which will yield precise intermediate frames by linearly combining bidirectional information. MPVF \cite{wei2023mpvf} resolves various magnitudes of motions by proposing a multi-pyramid voxel flows model that takes multi-scale voxel flows into account. DDM \cite{kim2022diffusion} can
learn spatial deformation information between the source and target volumes and provide a latent code for generating intermediate frames along
a geodesic path. UVI-Net \cite{kim2024data} utilizes the flow calculation model with the time-domain cycle-consistency constraint and linear motion hypothesis, to realize motion modeling in an unsupervised style. However, the linear hypothesis for motions will induce temporal inconsistency and spatial distortion.


\vspace{1mm}
\noindent \textbf{Diffusion-based Video Interpolation.}
Denoising diffusion probabilistic models \cite{ho2020denoising, ho2022video, liu2024sora} reveal promising performance in generating realistic images or videos. Some methods maximize the potential of diffusion models to resolve video interpolation tasks. Essentially, these approaches implicitly learn the temporal motions from the starting frame to the ending frame \cite{jain2024video, chen2024ultrasound, bae2024conditional}. LDMVFI \cite{danier2024ldmvfi} proposes a latent diffusion model, which approaches the VFI problem from a generative perspective by formulating it as a conditional synthesis issue. However, these diffusion models do not leverage effective conditional guidance, which is crucial for medical VFI with intrinsic motion priors. Thus, we introduce Fourier bases to enhance the conditional synthesis.




\section{Methodology}
\label{sec:formatting}

\subsection{Fourier Motion Bases}
Since respiratory motions for heart or lung exhibit a quasi-periodic characteristic, we intend to resolve this task from a frequency perspective. Specifically, the respiration can be viewed as an oscillatory process with certain frequencies. Thus, we draw an analogy between 2D time-varying curves and 4D temporal medical volumes. Curves can be decomposed into a group of harmonic oscillators with various frequencies \cite{tolstov2012fourier}. Similarly, temporal volumes can be restored via a stack of extracted Fourier bases, which are embodied with various motion patterns of anatomical deformation. The whole process can be formulated as:
\vspace{-3mm}
\begin{eqnarray}
    \mathcal{B}_k = \sum_{n=0}^{N-1} V_n e^{-i 2 \pi k \tfrac{n}{N}}, \, \, k = 0, 1, . . ., N-1, \\
    V_n = \frac{1}{T} \sum_{k=0}^{N-1} \mathcal{B}_k e^{i 2 \pi n \tfrac{k}{N}}, \, \, n = 0, 1, . . ., N-1,
    \label{FFT}
\end{eqnarray}
where $\mathcal{B}_{k}$ means the $k_{th}$ Fourier basis in the frequency domain, $V_{n}$ refers to the $n_{th}$ medical volume, and $N$ is the frame number. These Fourier bases describe locally anisotropic anatomical variations with different motion intensities, promoting the performance of the temporal interpolation model $\mathcal{I}$. Thus, the goal of our work is to deduce 4D medical videos $\bm{V}$ conditioned on two prompting frames $V_{0}$, $V_{N-1}$ and Fourier motion bases $\bm{\mathcal{B}}$:
\begin{eqnarray}
    & \mathcal{I}^{*} = \mathop{\arg\min} \limits_{\mathcal{I}} \, \| \, \bm{V} - \mathcal{I}(V_{0}, V_{N-1}; \bm{\mathcal{B}}) \,\|_{\mathcal{D}} \, ,
\end{eqnarray}
$\mathcal{I}^{*}$ refers to the optimal solution in the interpolation model set, and $\mathcal{D}$ represents the distance-based normed space. We systematically organize the following sections along Fourier bases. Section \ref{sec:FMO} introduces the design of the Fourier Motion Operator, which constructs Fourier bases with physiology motion priors and case-specific frequency information. Section \ref{sec:basis synthesis} adapts Fourier bases to the to-be-interpolated video domain with well-learned physiology motion priors. Section \ref{sec:4d diffusion} provides details of the diffusion model conditioned on Fourier bases.

\begin{figure}[!t]
\centerline{\includegraphics[width=0.95\linewidth]{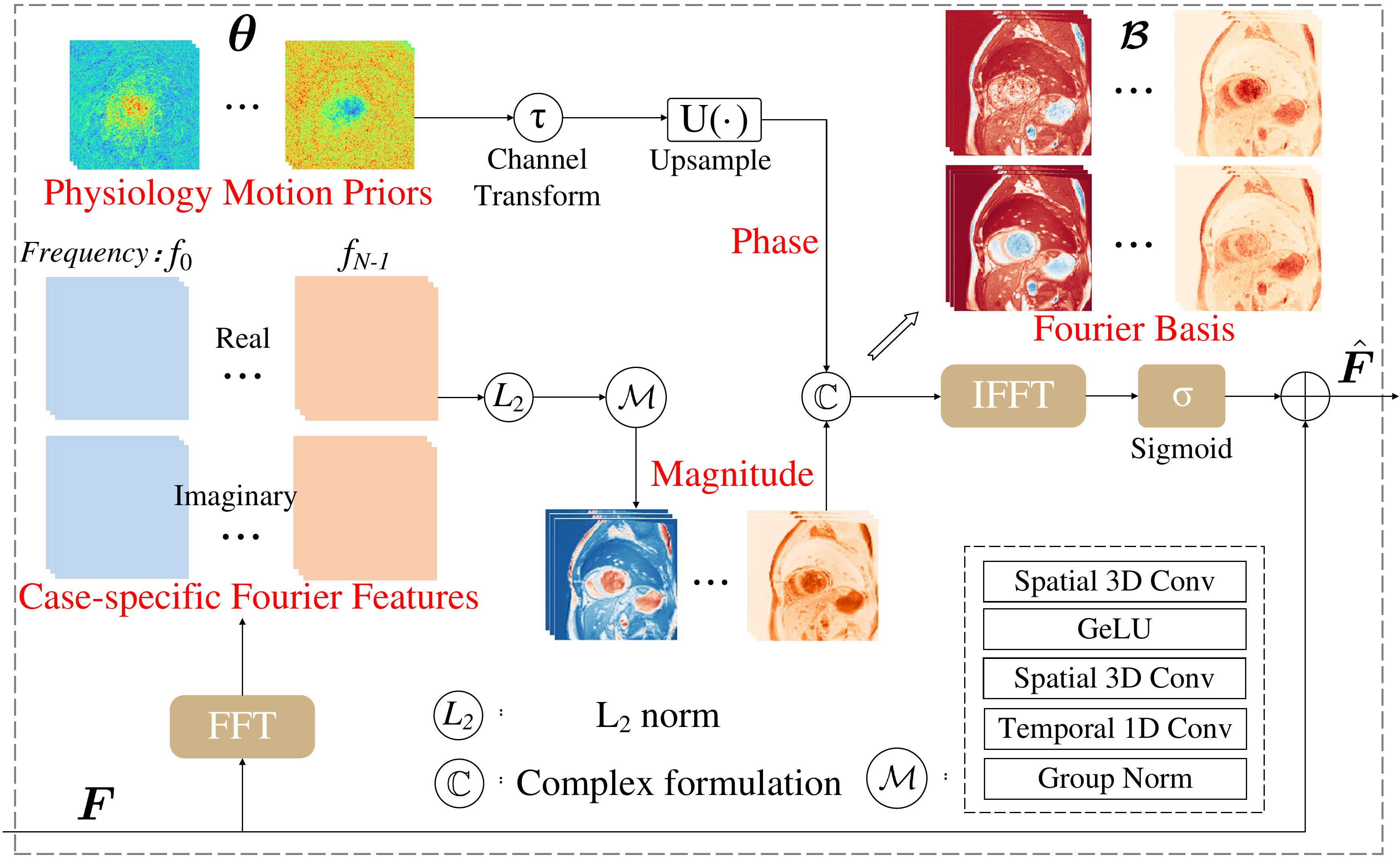}}
\caption{The details of Fourier motion operator. Fourier bases are decomposed as physiology motion priors and case-specific Fourier features, corresponding to phase and magnitude components.}
\label{FMO}
\end{figure}
\subsection{Fourier Motion Operator}
\label{sec:FMO}
To explicitly extract the Fourier bases with various motion patterns for the respiration process, we introduce the Fourier Operator $\mathcal{O}_{\mathcal{F}}$ as a neural operator. As stated in Eq.\eqref{FFT}, $\mathcal{O}_{\mathcal{F}}$ will operate in the temporal domain, not the spatial domain. Since the common Fourier operator $\mathcal{O}_{\mathcal{F}}$ only mines case-specific frequency representations, general physiology characteristics inside respiratory data are ignored. Inspired by that notion, we devise the Fourier motion operator $\mathcal{O}_{FM}$ as a neural operator in the feature space. To map the image space into feature space, we introduce a Variational Autoencoder (VAE) as revealed in Fig.\ref{pipeline}(a). Due to no built-in 4D convolutional operator, we replace it with 3D spatial convolution and 1D temporal convolution as \cite{liang2024diffusion4d, zhang20254diffusion} did. The spatial layer in Fig.\ref{pipeline}(a) consists of spatial convolution, GELU, and group normalization operations.

The Fourier motion operator $\mathcal{O}_{FM}$ is devised to acquire spectral characteristics with both physiology motion priors and case-specific representations in the feature space. Firstly, physiology motion priors are introduced as learnable frequency embeddings, which are updated by backward gradients of restoring given temporal data via VAE. Well-trained frequency embeddings $\bm{\theta}$ can represent the generic discipline of the regular respiration process \cite{you2025slord}, with $\theta_{k}$ referring to the $k_{th}$ element ($k=0,1,...,N-1$). Then Fourier bases $\bm{\mathcal{B}}$ are made up of physiology motion priors $\bm{\theta}$ and case-specific features $\bm{F}$, which arise from the output of spatial and temporal layers. As depicted by Fig.\ref{FMO}, $\bm{\mathcal{B}}$ with a complex type is yielded in the following style:
\begin{eqnarray}
    & \bm{\mathcal{B}} = \underbrace{\mathcal{M}(\|\mathcal{O}_{\mathcal{F}}(\bm{F}))\|_{2})}_{\mathrm{Case\mbox{-}Specific}} \cdot \underbrace{e^{i \,U(\tau(\bm{\theta}))}}_{\mathrm{General}} \, \, ,
\end{eqnarray}
$\tau$ means the channel transform, $U$ refers to the upsampling which unifies the resolution of $\bm{\theta}$ and $\bm{F}$. These operations transform general motion priors into the phase part of Fourier bases, which determines the motion variation standing for the whole dataset distribution. Then $\|\cdot\|_{2}$ means the $L_{2}$ norm, $\mathcal{M}$ is the convolutional mapping operation. They will output motion bases' magnitude information, which selectively highlights regions in spectral maps relevant to case-specific information.

Meanwhile, feature maps $F$ are further enriched with more fine-grained frequency details. The specific mechanism can be described with Eq.\eqref{feature enhance}:
\begin{eqnarray}
    & \hat{\bm{F}} = \bm{F} + \sigma\mathcal{O}_{\mathcal{F}}^{-1}(\bm{\mathcal{B}}),
    \label{feature enhance}
\end{eqnarray}
here $\mathcal{O}_{\mathcal{F}}^{-1}$ and $\sigma$ refer to the inverse Fourier transform and Sigmoid operators, which will boost deep features $\hat{\bm{F}}$ with enriched details, including shape and texture information.

\subsection{Frame Inpainting-based Basis Synthesis}
\label{sec:basis synthesis}
Due to the fact that only starting and ending frames can be acquired during inference, a significant step is to adapt Fourier bases to the masked video domain. Motivated by the fact that strong correlations exist between Fourier motion bases and inpainted temporal sequences, we plan to jointly optimize these two variables. To accomplish that goal, we repurpose well-calibrated physiology motion priors from VAE, which prove to be effective in providing the motion guidance for intermediate frame inpainting. Also, pretrained weights of VAE encode implicit temporal dependencies between frames, enhancing the reconstruction of masked video sequences via the finetuned VAE noted as $\mathcal{G}$. A mathematical formulation is defined as:
\begin{eqnarray}
    & \hat{\bm{V}}, \hat{\bm{\mathcal{B}}} = \mathcal{G} (\hat{\bm{\theta}}, \bm{V_{M}}) = \mathcal{G} (\hat{\bm{\theta}}, V_{0}, .., \bm{0}, .., V_{N-1}),
\end{eqnarray}
$\hat{\bm{\theta}}$ and $\bm{V_{M}}$ correspond to well-trained general motions and zero-masked videos, $\hat{\bm{V}}$ and $\hat{\bm{\mathcal{B}}}$ refer to inpainted videos and domain-adapted Fourier bases.

The interaction between temporal video representations and $\hat{\bm{\mathcal{B}}}$ is implemented via the Fourier motion operator as shown in Fig.\ref{pipeline}(b). For the joint optimization process, video sequences will contribute temporal variations, enhancing the synthesis of adapted Fourier bases. Conversely, Fourier bases for zero-masked inputs inherently generate a spectral map that emphasizes regions with motion in the temporal axis. This activated map aids in temporal frame inpainting by exerting attention on regions with different magnitudes.

\begin{figure}[!t]
\centerline{\includegraphics[width=0.95\linewidth]{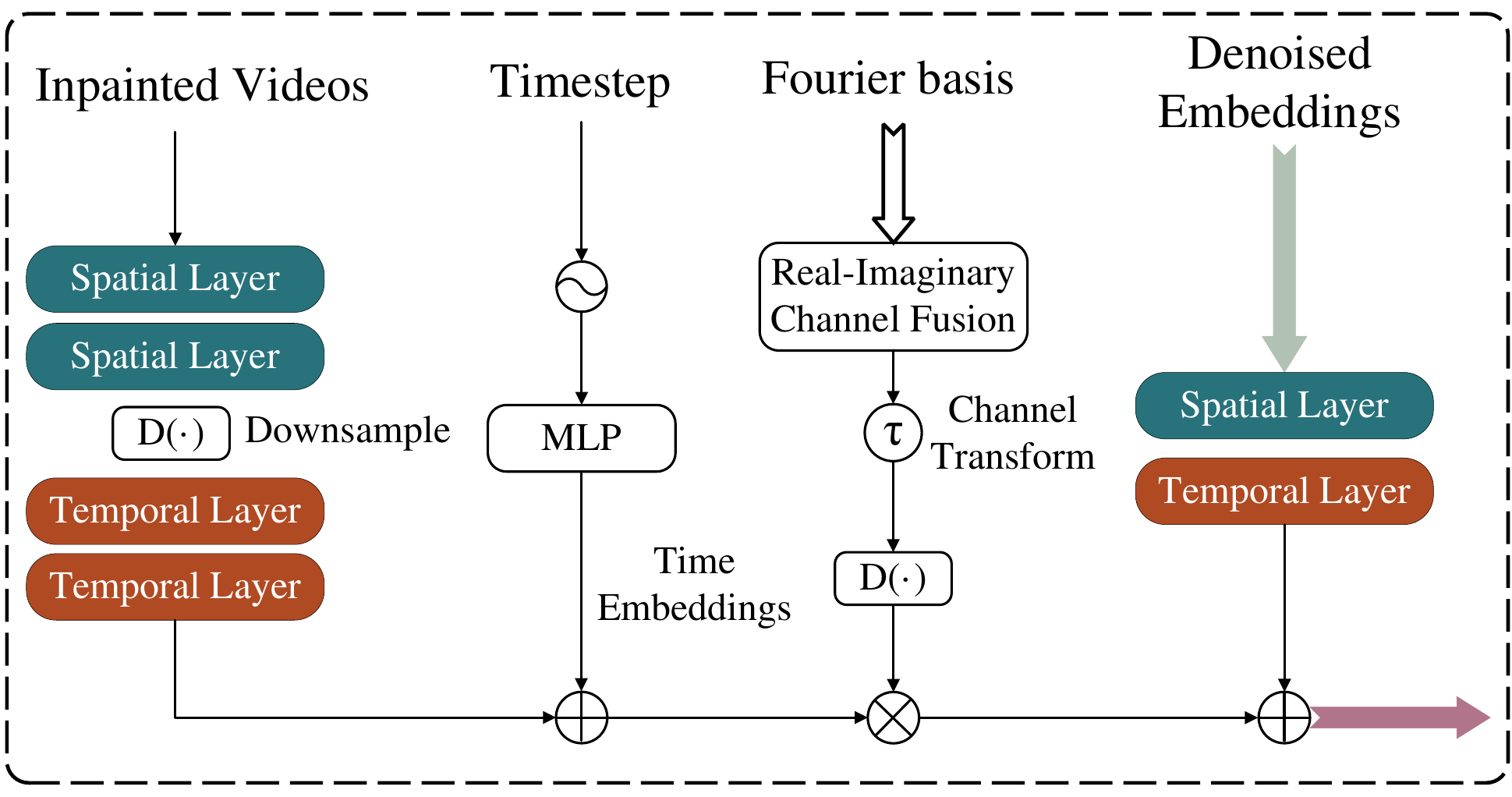}}
\caption{The architecture of basis interaction operator. Fourier
motion bases selectively enhance video representations by measuring the motion intensity of different regions. Then those enhanced temporal representations can enrich denoised embeddings from the diffusion model.}
\label{BIO}
\end{figure}

\subsection{Basis-guided 4D Conditional Diffusion}
\label{sec:4d diffusion}

VAE in Fig.\ref{pipeline}(a) is not only devised for universal motion priors, but also for the latent diffusion model (LDM), which shows better compatibility for extra conditional inputs. Besides, LDM synthesizes latent embeddings occupying less GPU memory compared to 4D temporal volumes. And the VAE decoder is leveraged to reconstruct interpolated video sequences with synthesized latent embeddings.

\begin{table*}[!t]
  \begin{center}
    \caption{Baseline comparison with flow-based and diffusion-based models. (Bold: the best, Underlined: the second best. $\Delta$ values represent the relative difference between our model and UVI-Net, and \textcolor{red}{red} and \textcolor{blue}{blue} numbers stand for better and worse results).}
\label{benchmark_table}
    \centering
    \resizebox{0.90\textwidth}{!}{
    \begin{tabular}{ccccccccc}
        \hline
   \multirow{2}*{Model} & \multicolumn{4}{c}{ACDC Cardiac } & \multicolumn{4}{c}{4D Lung}  \\  
  \cmidrule(r){2-5}    \cmidrule(r){6-9}
    & PSNR (dB) $\uparrow$ & LPIPS $\downarrow$ & FID $\downarrow$ &  FVD $\downarrow$ & PSNR (dB) $\uparrow$ & LPIPS $\downarrow$ & FID $\downarrow$ & FVD $\downarrow$ \\
            \hline
        SVIN \cite{guo2020spatiotemporal} & 31.43\scriptsize{$\pm0.421$} & \underline{1.563\scriptsize{$\pm0.206$}} & 26.3 &  93.6 & 30.49\scriptsize{$\pm0.304$} & 2.650\scriptsize{$\pm0.245$} & 38.5 & 125.6 \\ 
        Voxelmorph \cite{balakrishnan2019voxelmorph} & 30.77\scriptsize{$\pm0.502$} & 1.969\scriptsize{$\pm0.197$}  & 29.4 & 102.1 & 29.90\scriptsize{$\pm0.373$} & 2.815\scriptsize{$\pm0.260$} & 43.2 & 149.0 \\
        Transmorph \cite{chen2022transmorph} & 29.81\scriptsize{$\pm0.532$} & 2.136\scriptsize{$\pm0.255$}  & 31.5 & 109.5 & 29.04\scriptsize{$\pm0.403$} & 2.946\scriptsize{$\pm0.291$} & 40.5 & 143.5 \\
        MPVF \cite{wei2023mpvf}  & \underline{31.53\scriptsize{$\pm0.410$}} & 1.594\scriptsize{$\pm0.201$}  & 24.8 & \underline{91.3} & \underline{30.76\scriptsize{$\pm0.265$}} & 2.418\scriptsize{$\pm0.246$} &  39.4 & \underline{120.3} \\
     UVI-Net \cite{kim2024data} & \textbf{32.16\scriptsize{$\pm0.402$}} & 1.662\scriptsize{$\pm0.245$} & \underline{23.0} &  94.2 & \textbf{31.57\scriptsize{$\pm0.311$}} & \underline{2.211\scriptsize{$\pm0.216$}} & \underline{37.8} & 121.7  \\ 
     IFRNet \cite{kong2022ifrnet}  & 30.58\scriptsize{$\pm0.408$} & 1.752\scriptsize{$\pm0.249$}  & 27.8 & 97.5 & 30.05\scriptsize{$\pm0.280$} & 2.316\scriptsize{$\pm0.273$} &  39.7 & 124.0 \\
        LDMVFI \cite{danier2024ldmvfi}  & 27.11\scriptsize{$\pm0.460$} & 2.943\scriptsize{$\pm0.410$} & 28.8 & 99.2 & 26.31\scriptsize{$\pm0.453$} & 3.659\scriptsize{$\pm0.341$} & 42.1 & 142.7 \\
        LDDM \cite{chen2024ultrasound} & 24.53\scriptsize{$\pm0.481$} & 2.634\scriptsize{$\pm0.308$} & 33.4 & 105.7 & 25.19\scriptsize{$\pm0.273$} & 2.914\scriptsize{$\pm0.287$} & 48.7 &  146.3 \\
      Conditional Diff \cite{ho2022video} & 26.59\scriptsize{$\pm0.545$} & 2.460\scriptsize{$\pm0.357$} & 30.1 &  95.7 & 25.95\scriptsize{$\pm0.391$} & 3.294\scriptsize{$\pm0.375$} & 45.0 &  133.0 \\ 
        DDM \cite{kim2022diffusion} & 29.79\scriptsize{$\pm0.504$} & 2.089\scriptsize{$\pm0.352$} & 27.4 & 110.3 & 29.67\scriptsize{$\pm0.420$} & 2.705\scriptsize{$\pm0.330$} & 40.8 & 165.5 \\
                    \hline
        FB-Diff  & 30.95\scriptsize{$\pm0.386$} & \textbf{1.452\scriptsize{$\pm0.205$}} & \textbf{21.5} & \textbf{88.9} & 30.18\scriptsize{$\pm0.287$} & \textbf{2.066\scriptsize{$\pm0.215$}} & \textbf{37.7} & 
   113.6 \\
   $\Delta$ value  & \textcolor{blue}{1.21 $\downarrow$} & \textcolor{red}{0.210 $\downarrow$} & \textcolor{red}{1.5 $\downarrow$} & \textcolor{red}{5.3 $\downarrow$} & \textcolor{blue}{1.39 $\downarrow$} & \textcolor{red}{0.145 $\downarrow$} & \textcolor{red}{0.1 $\downarrow$} & \textcolor{red}{8.1 $\downarrow$} \\
        \hline 

    \end{tabular}}
  \end{center}
\end{table*}

For the specific diffusion process, the whole video sequence $\bm{V}$ is first encoded as a latent embedding $z_{0}$ = $\mathcal{G}_{e}(\bm{V})$. Then the latent code $z_{0}$ is perturbed as:
\begin{eqnarray}
  & z_{t} = \sqrt{\overline{\alpha}_{t}} \; z_{0} + \sqrt{1 - \overline{\alpha}_{t}} \; \epsilon, \, \epsilon \sim \bm{\mathcal{N}}(0, 1),
\end{eqnarray}
where $\overline{\alpha}_{t} = \prod_{i=1}^{t} (1 - \beta_{t}) $ with $\beta_{t}$ is the noise coefficient
at time step $t$, and $t$ is uniformly sampled from the timestep index set
$[1, 2, . . . , T]$. This process can be regarded as a Markov chain, which
incrementally adds Gaussian noise to the latent code $z_{0}$. The denoising model $d$ receives $z_{t}$ as input and is optimized under conditional guidance to learn the
latent space distribution with the objective function:
\begin{eqnarray}
  & \mathcal{L}_{\epsilon} = E_{z_{t}, \epsilon \sim \mathcal{N}(0, 1)} \; \| \epsilon - \epsilon_{d} (z_{t}, t; \hat{\bm{\mathcal{B}}}, \mathcal{G}_{e}(\bm{V_{M}}), \hat{\bm{V}}) \|^{2}.
\label{diffusion loss}
\end{eqnarray}
here $\epsilon_{d}$ is the regressed noise, adapted Fourier bases $\hat{\bm{\mathcal{B}}}$ and inpainted coarse videos $\hat{\bm{V}}$ are incorporated as the conditional guidance. Moreover, feature maps of starting and ending frames by the VAE encoder $\mathcal{G}_{e}$ serve as a strong prompt for the 4D video synthesis. Then during the training process of FB-Diff, the total loss $\mathcal{L}$ consists of the denoising loss $\mathcal{L}_{\epsilon}$ and the $L_{2}$-norm reconstruction loss $\mathcal{L}_{r}(\hat{\bm{V}}, \bm{V})$.

To inject Fourier motion bases into LDM and harness coarsely inpainted results, we design the Basis Interaction Operator (BIO) as shown in Fig.\ref{BIO}. Specifically, inpainted videos are mapped into video feature representations via spatial \& temporal layers and the downsampling operation. Fourier motion bases with various motion patterns, selectively enhance video representations by measuring the motion intensity of different regions. Those enhanced temporal representations can enrich denoised embeddings from the diffusion model, promoting a promising synthesis of highlighted regions. Thus, BIO can boost well-interpolated videos with spatial and temporal consistency.



\section{Experiments}
\subsection{Experimental Settings}
\textbf{Dataset.} To evaluate the interpolation performance of our framework, we conduct experiments on public ACDC cardiac \cite{bernard2018deep} and 4D-Lung datasets \cite{hugo2016data}. For ACDC with the MRI modality, volume sequences between the end-diastolic and end-systolic phases are extracted as valid 4D data. Of all the 150 4D sequences, cases with identity $1\mbox{-}100$, $101\mbox{-}120$, and $121\mbox{-}150$ serve as the training, validation, and testing sets. Besides, all MRI volumes are cropped to $128 \times 128 \times 32$, and the frame number ranges from $6$ to $16$ \cite{you2024learning, zhang2024pass}. For 4D-Lung CT images, the end-inspiratory and end-expiratory scans are set as the initial and final images. We collect 125 4D videos from \cite{hugo2016data}, with the resolution equal to $6 \times 128 \times 128 \times 128$. They are split as $80/15/30$ cases for training, validation, and inference. We uniform the frame number as 16 with zero paddings.

\noindent \textbf{Evaluations.} We select quantitative evaluation metrics including Peak Signal-to-Noise Ratio (PSNR) \cite{hore2010image}, Learned Perceptual Image Patch Similarity (LPIPS) \cite{zhang2018unreasonable, you2025temporal}, Fréchet Inception Distance (FID) \cite{heusel2017gans}, and Fréchet Video Distance (FVD) \cite{unterthiner2018towards}. Compared to PSNR, which is the traditional pixel-wise reconstruction metric, LPIPS and FVD evaluate perceptual quality from the perspective of human vision \cite{zhong2024clearer, jain2024video, danier2024ldmvfi}, which is more feasible and favored in clinical practice. Specifically, LPIPS reveals important perceptual similarity like human vision. FID and FVD evaluate the dataset distribution bias between realistic and synthesized data, including spatial and temporal consistency.

\noindent \textbf{Training Details.} The VAE maps the image space into the downsampled latent space with a ratio of $1/8$. And the proposed conditional diffusion model is implemented with a lightweight UNet. All models are trained using the AdamW optimizer with the linear warm-up strategy. The batch size, diffusion step $T$, and training epoch are equal to $2$, $1000$, and $500$. Experiments are implemented based on Pytorch and 2 NVIDIA RTX 4090 GPUs. More details about the network and training settings are described in the Appendix. 
\begin{figure*}[!t]
\centerline{\includegraphics[width=0.97\linewidth]{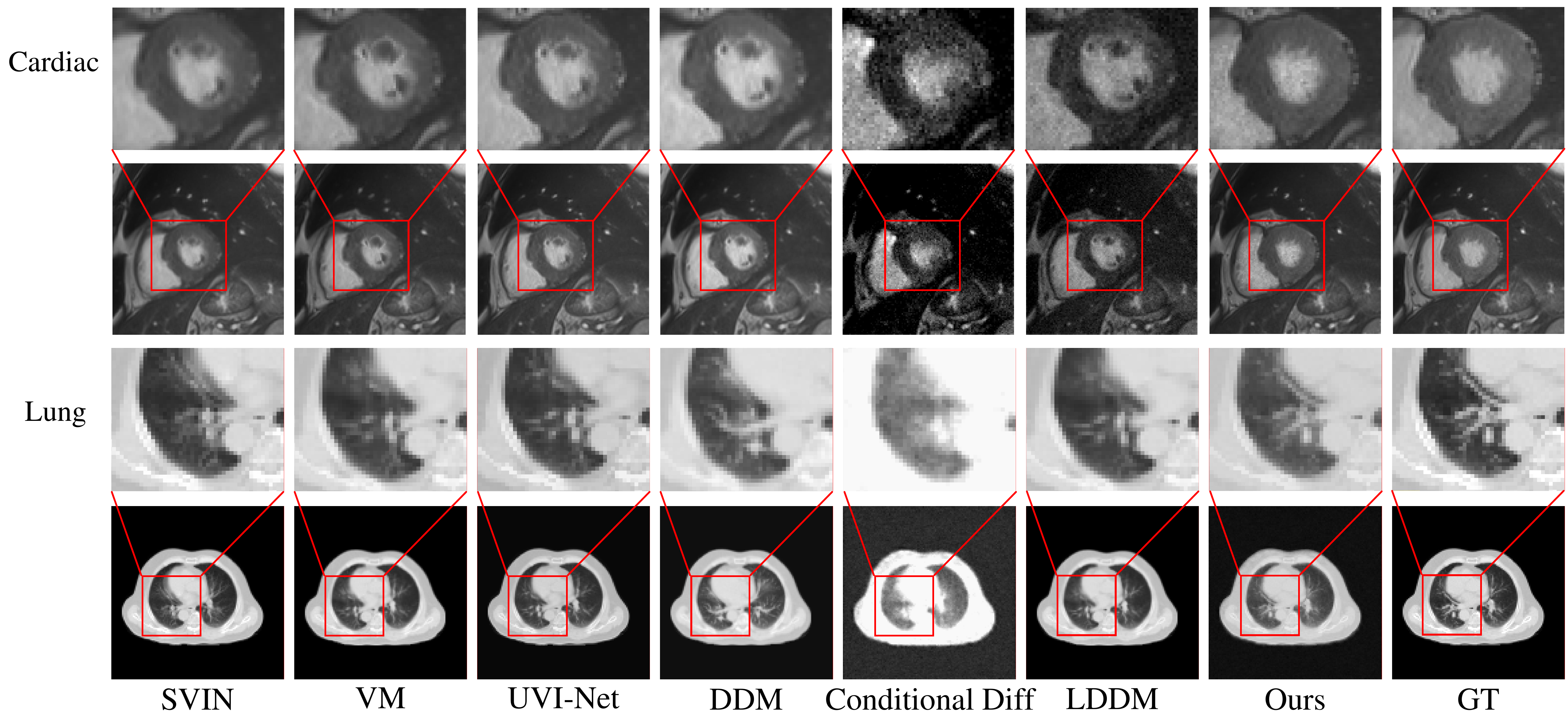}}
\caption{Qualitative comparison of the midmost frame interpolation between different models. VM: Voxelmorph.}
\label{FIGURE_BENCHMARK}
\end{figure*}

\subsection{Comparisons with SOTA baselines}
We compare the proposed FB-Diff with ten classical baselines, including flow-based methods with the linear motion hypothesis \cite{guo2020spatiotemporal, balakrishnan2019voxelmorph, chen2022transmorph, wei2023mpvf, kim2024data, kong2022ifrnet} and diffusion-based models \cite{danier2024ldmvfi, chen2024ultrasound, kim2022diffusion, ho2022video}.

\noindent \textbf{Quantitative Comparison.} We comprehensively conduct experiments on ACDC cardiac and 4D Lung datasets. Quantitative results are illustrated in Table \ref{benchmark_table}. Our model achieves the best performance across the benchmark on perceptual evaluation metrics from the perspective of human vision \cite{zhong2024clearer, danier2024ldmvfi}, which are clinically favored compared with the pixel-centric reconstruction metric \cite{chen2024ultrasound}. Specifically, FB-Diff outperforms UVI-Net \cite{kim2024data} with $0.210\downarrow$ LPIPS and $5.3\downarrow$ FVD values on ACDC. Also, our model maintains quite satisfactory PSNR values of $30.95$ dB and $30.18$ dB on ACDC and 4D Lung. Although diffusion models are not sensitive to pixel-wise intensities \cite{ho2020denoising}, the conditional guidance of Fourier bases and coarsely inpainted videos is effective to boost the interpolation performance of diffusion models. Table \ref{benchmark_table} reveals that FB-Diff significantly surpasses existing diffusion models on all metrics.

Furthermore, the interpolation process becomes most challenging for the midmost frames since they are positioned farthest from the two reference frames \cite{jain2024video}. And our model can theoretically resolve this issue due to stronger temporal representations based on the nonlinear and quasi-periodic motion modeling. Thus, a quantitative evaluation of central frames is necessary. As shown in Table \ref{midmost_frame}, FB-Diff demonstrates even greater performance advantages, with $4.3\downarrow$ FID and $8.2\downarrow$ FVD values compared with UVI-Net. Additionally, FB-Diff acquires a competitive enough PSNR value, which demonstrates the effectiveness of respiratory modeling via Fourier bases.

\noindent \textbf{Qualitative Comparison.} Qualitative results on the interpolated midmost frame are depicted in Fig.\ref{FIGURE_BENCHMARK}. SVIN \cite{guo2020spatiotemporal}, Voxelmorph \cite{balakrishnan2019voxelmorph}, and DDM \cite{kim2022diffusion} leverage the single-direction flow estimation to interpolate intermediate frames with linear motions. Compared with that, UVI-Net \cite{kim2024data} constructs the interpolation model with time-domain cycle-consistency, and achieves better reconstruction performance as expected. However, these methods essentially overlook the nonlinear and quasi-periodic motion discipline between temporal sequences, resulting in unsatisfactory interpolated results. Instead, FB-Diff exhibits advantageous predictions for midmost frames under the guidance of various motion patterns, indicating locally anisotropic motions with different intensities. Besides, other diffusion models do not work well due to the lack of fine-grained conditional guidance.

\begin{table}[!t]
  \begin{center}
    \caption{Interpolation performance of the midmost frames on ACDC cardiac. The choice of midmost frames: if the frame number $N$ is odd, three intermediate frames are selected, otherwise only two frames are chosen.}
\label{midmost_frame}
    \centering
    \resizebox{1.0\columnwidth}{!}{
    \begin{tabular}{ccccc}
        \hline
    Model & PSNR (dB) $\uparrow$ & LPIPS $\downarrow$ & FID $\downarrow$ &  FVD $\downarrow$  \\
            \hline
        SVIN \cite{guo2020spatiotemporal} & 26.01\scriptsize{$\pm0.467$} & 2.673\scriptsize{$\pm0.361$} & 38.9 &  142.8 \\ 
        DDM \cite{kim2022diffusion} & 24.37\scriptsize{$\pm0.511$} & 2.756\scriptsize{$\pm0.330$} & 41.5 &  145.1 \\ 
        UVI-Net \cite{kim2024data} & 26.25\scriptsize{$\pm0.410$} & 2.542\scriptsize{$\pm0.374$} & 37.0 &  137.1 \\ 
        \hline
        FB-Diff  & 26.18\scriptsize{$\pm0.413$} & 2.287\scriptsize{$\pm0.305$} & 32.7 & 128.9 \\
       $\Delta$ value  & \textcolor{blue}{0.07 $\downarrow$} & \textcolor{red}{0.255 $\downarrow$} & \textcolor{red}{4.3 $\downarrow$} & \textcolor{red}{8.2 $\downarrow$} \\
        \hline 
    \end{tabular}}
  \end{center}
\end{table} 

\begin{figure*}[!t]
\centerline{\includegraphics[width=1.0\linewidth]{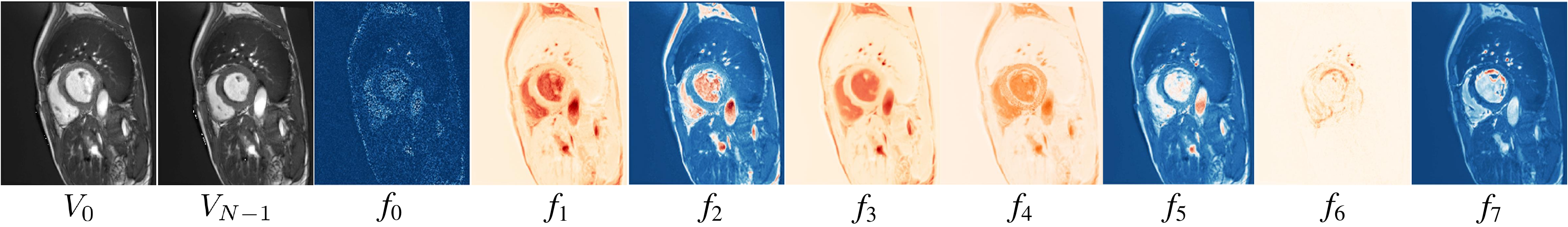}}
\caption{Visualizations on the spectral intensity of Fourier bases. $V_{0}$, $V_{N-1}$: starting and ending frames. $f_{1}$-$f_{8}$ refer to different motion patterns, which highlight different regions under different frequency. There also appear locally anisotropic motion intensities inside a specific motion pattern.}
\label{motion_visualize}
\end{figure*}

\begin{table}[!t]
  \begin{center}
  \caption{Ablation study on key components. $\bm{\theta}$: physiology motion priors, $\hat{\bm{\mathcal{B}}}$: Fourier motion bases,  $\hat{\bm{V}}$: coarsely inpainted videos.}
\label{ablation_table}
  \resizebox{1.0\columnwidth}{!}{
  \begin{tabular}{ccccccc}  
  \hline  
  \multicolumn{3}{c}{Settings}  & \multicolumn{2}{c}{ACDC Cardiac} & \multicolumn{2}{c}{4D Lung}  \\  
    \cmidrule(r){1-3}    \cmidrule(r){4-5}  \cmidrule(r){6-7}
    $\bm{\theta}$  & $\hat{\bm{\mathcal{B}}}$   & $\hat{\bm{V}}$  & PSNR (dB) $\uparrow$  & LPIPS $\downarrow$  & PSNR (dB) $\uparrow$ & LPIPS  $\downarrow$ \\
  \hline   
   \textcolor{red}{\ding{55}}  & \textcolor{green}{\ding{52}} & \textcolor{green}{\ding{52}}   & 29.76 & 1.710  & 29.53 & 2.413   \\
  \textcolor{red}{\ding{55}} & \textcolor{red}{\ding{55}} & \textcolor{green}{\ding{52}}  & 28.17 & 2.105  & 28.40 & 2.971  \\
  \textcolor{green}{\ding{52}} & \textcolor{green}{\ding{52}} &  \textcolor{red}{\ding{55}}  & 30.29 & 1.655  & 29.71 & 2.246 \\
   \textcolor{green}{\ding{52}} & \textcolor{green}{\ding{52}} &  \textcolor{green}{\ding{52}} & 30.95 & 1.452  & 30.18 & 2.066 \\
  \hline
  \end{tabular}}
  \end{center}
\end{table}

\subsection{Ablation Study}
\noindent \textbf{Key Components.} An ablation study is conducted to investigate the significance of various components in FB-Diff as shown in Table \ref{ablation_table}. (a) \textbf{Physiology Motion Priors} $\bm{\theta}$. By removing $\bm{\theta}$ from the Fourier motion operator, interpolation performance decreases significantly due to the loss of general spectral characteristics from respiratory data. Therefore, Fourier bases lack the precise phase information. (b) \textbf{Fourier bases} $\hat{\bm{\mathcal{B}}}$. Without $\hat{\bm{\mathcal{B}}}$ as conditional guidance, the conditional diffusion model fails to perceive explicit motion patterns with different frequencies, which provide the intensity of anisotropic motions. As a result, the generative model cannot well learn the regular temporal variations, with $2.78$ dB $\downarrow$ and $1.78$ dB $\downarrow$ PSNR values on ACDC and 4D Lung. (c) \textbf{Inpainted Videos} $\hat{\bm{V}}$. Here the coarsely reconstructed sequences serve as the output by the finedtuned VAE as revealed in Fig.\ref{pipeline}. And the joint optimization process boosts the spatial and temporal representations of $\hat{\bm{V}}$. Consequently, discarding this component will reduce fine-grained details of the diffusion model's output.

\noindent \textbf{Hyper-settings of Fourier Bases.} (a) \textbf{The number of Fourier Bases} $N_{Fb}$. We first analyze the quantitative effects of the frequency number on the interpolation performance. As revealed in Table \ref{ablation_}, reducing the frequency number will cause degraded performance of the temporal simulation. Specifically, $1.30$ dB lower PSNR and $0.535$ higher LPIPS values are induced by decreasing $N_{Fb}$ from $16$ to $8$. That phenomenon indicates that some elements in the group of bases with the frequency ranging from $f_{8}$ to $f_{15}$ are consistent with specific frequencies of the respiratory process. (b) \textbf{Real \& Imaginary Parts}. $\hat{\bm{\mathcal{B}}}$ bears real and imaginary parts after combining generic priors and case-specific spectral representations, which perform as the phase and magnitude information. Eliminating either the real or imaginary part will convert the complex data type into the real one, resulting in the loss of phase information and influencing the temporal interpolation.

\begin{table}[!t]
  \begin{center}
  \caption{Ablation study on the hyper-settings of Fourier bases. $N_{Fb}$: the frequency number of Fourier bases.}
\label{ablation_}
  \resizebox{1.0\columnwidth}{!}{
  \begin{tabular}{lcccc}  
  \hline  
  \multirow{2}*{Settings}  & \multicolumn{2}{c}{ACDC Cardiac} & \multicolumn{2}{c}{4D Lung}  \\  
    \cmidrule(r){2-3}    \cmidrule(r){4-5}
    & PSNR (dB) $\uparrow$  & LPIPS $\downarrow$  & PSNR (dB) $\uparrow$ & LPIPS  $\downarrow$ \\
  \hline   
  $N_{Fb}=16$ (Ours) & 30.95 & 1.452  & 30.18 & 2.066 \\
  $N_{Fb}=12$ & 30.76 & 1.640  & 29.97 & 2.570   \\
  $N_{Fb}=8$ & 29.65 & 1.987 & 29.50 & 2.745 \\
\hline
  w/o Real part & 29.18 & 1.872  & 29.56 & 2.453   \\
  w/o Imaginary part & 29.88 & 1.681 & 29.67 & 2.401 \\
\hline
  \end{tabular}}
  \end{center}
\end{table}

\subsection{Visualization Analysis}
\noindent \textbf{Visualization of Fourier bases.} To further validate the effectiveness and interpretability of our proposed Fourier bases, we visualize extracted motion bases from the most shadow layer. As revealed in Fig.\ref{motion_visualize}, $f_{0}-f_{7}$ refer to the magnitude of the first eight motion bases in ascending order of frequency. Specifically, $f_{0}$ shows a low-frequency motion pattern highlighting no region, which is strongly related to static backgrounds. $f_{1}-f_{3}$ exhibit a large motion intensity in the highlighted right and left ventricle regions. Those regions are indeed dynamic foregrounds requiring special attention. $f_{4}$ depicts an activation in the myocardial region, which contracts inward along the circumferential direction. $f_{5}-f_{7}$ highlight regions with vast deformations in the temporal domain. Particularly, $f_{6}$ illustrates a large motion on boundaries of anatomical structures. In summary, these motion patterns reveal locally anisotropic motion intensities, and are harmonized with specific frequencies of the respiratory process, thus promoting the temporal interpolation.

\begin{figure}[!t]
\centerline{\includegraphics[width=0.95\linewidth]{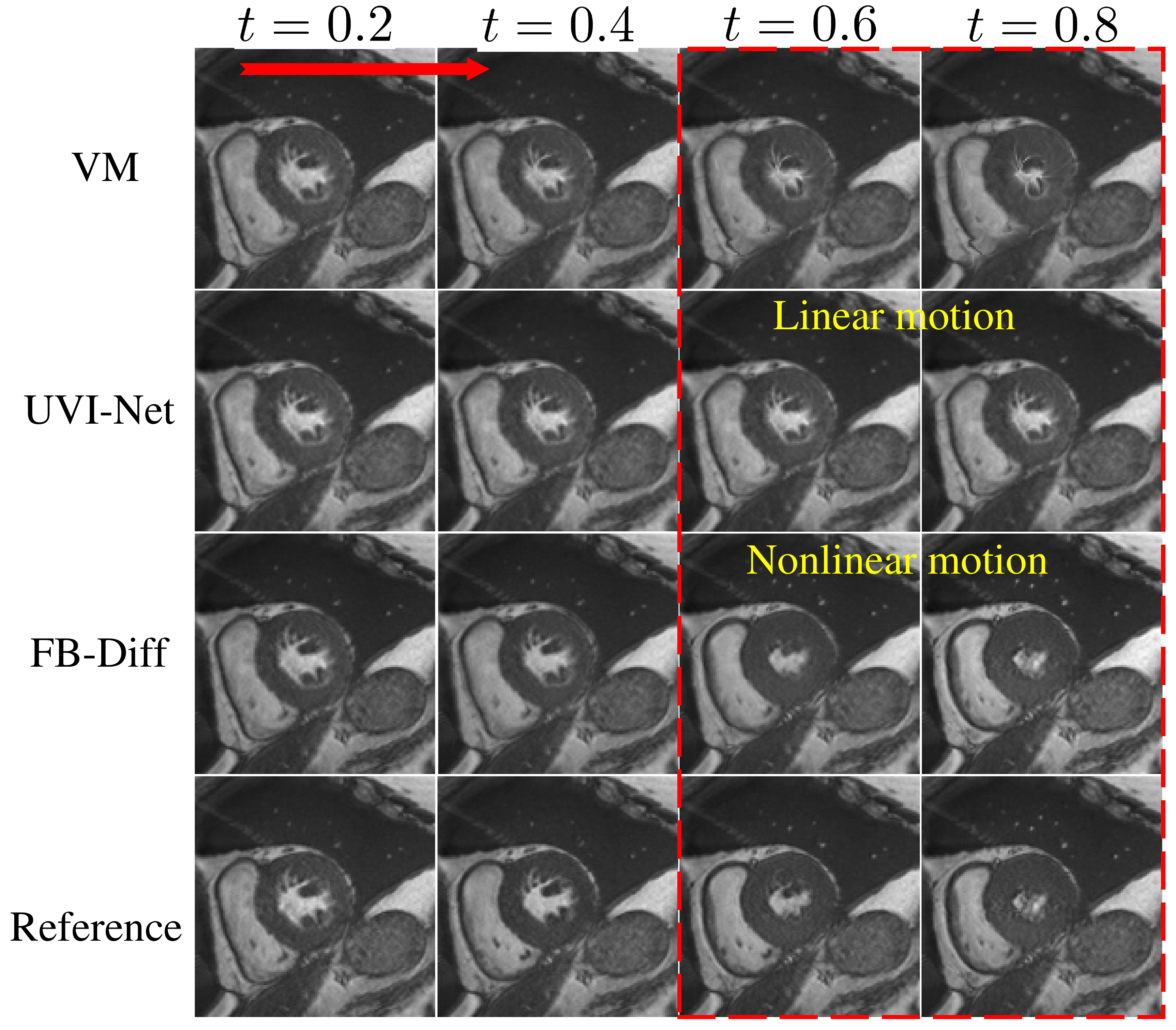}}
\caption{Temporal variation comparison between FB-Diff and existing methods with the linear motion hypothesis.}
\label{temporal variation}
\end{figure}
\noindent \textbf{Visualization of temporal anatomical variations.} To demonstrate the effectiveness of FB-Diff on temporal simulation, we visualize temporal variation maps for a perceptual comparison. Fig.\ref{temporal variation} provides interpolated intermediate frames at different timepoints. According to the referenced video, the anatomical motion in the temporal domain is nonlinear, which cannot be precisely simulated by existing methods with the linear motion hypothesis. In contrast, our predictions show better temporal consistency and continuity, particularly for frames at $t=0.6$ and $0.8$.

\section{Conclusion}
Our proposed FB-Diff resolves the temporal interpolation task of 4D medical imaging from a frequency perspective, and can model the respiratory process with nonlinear and quasi-periodic motions. A Fourier motion operator is elaborately devised to extract Fourier bases by incorporating physiological motion priors and case-specific spectral information in the feature space. And the diffusion model further leverages Fourier bases and two prompting frames to achieve the precise interpolation. Extensive results demonstrate that FB-Diff achieves SOTA perceptual performance with better temporal consistency while maintaining promising reconstruction metrics. Ablation studies demonstrate the efficacy of key components. Ultimately, it requires more conditional guidance such as electrocardiogram signals to simulate the highly unstable breathing, which is promising in future work. 

{
    \small
    \bibliographystyle{ieeenat_fullname}
    \bibliography{main}

\begin{thebibliography}{60}
\providecommand{\natexlab}[1]{#1}
\providecommand{\url}[1]{\texttt{#1}}
\expandafter\ifx\csname urlstyle\endcsname\relax
  \providecommand{\doi}[1]{doi: #1}\else
  \providecommand{\doi}{doi: \begingroup \urlstyle{rm}\Url}\fi

\bibitem[Bae et~al.(2024)Bae, Tong, and Chen]{bae2024conditional}
Juyoung Bae, Elizabeth Tong, and Hao Chen.
\newblock Conditional diffusion model for versatile temporal inpainting in 4d cerebral ct perfusion imaging.
\newblock In \emph{MICCAI}, pages 67--77. Springer, 2024.

\bibitem[Balakrishnan et~al.(2019)Balakrishnan, Zhao, Sabuncu, Guttag, and Dalca]{balakrishnan2019voxelmorph}
Guha Balakrishnan, Amy Zhao, Mert~R Sabuncu, John Guttag, and Adrian~V Dalca.
\newblock Voxelmorph: a learning framework for deformable medical image registration.
\newblock \emph{IEEE transactions on medical imaging}, 38\penalty0 (8):\penalty0 1788--1800, 2019.

\bibitem[Bao et~al.(2019)Bao, Lai, Zhang, Gao, and Yang]{bao2019memc}
Wenbo Bao, Wei-Sheng Lai, Xiaoyun Zhang, Zhiyong Gao, and Ming-Hsuan Yang.
\newblock Memc-net: Motion estimation and motion compensation driven neural network for video interpolation and enhancement.
\newblock \emph{IEEE transactions on pattern analysis and machine intelligence}, 43\penalty0 (3):\penalty0 933--948, 2019.

\bibitem[Bernard et~al.(2018)Bernard, Lalande, Zotti, Cervenansky, Yang, Heng, Cetin, Lekadir, Camara, Ballester, et~al.]{bernard2018deep}
Olivier Bernard, Alain Lalande, Clement Zotti, Frederick Cervenansky, Xin Yang, Pheng-Ann Heng, Irem Cetin, Karim Lekadir, Oscar Camara, Miguel Angel~Gonzalez Ballester, et~al.
\newblock Deep learning techniques for automatic mri cardiac multi-structures segmentation and diagnosis: is the problem solved?
\newblock \emph{IEEE transactions on medical imaging}, 37\penalty0 (11):\penalty0 2514--2525, 2018.

\bibitem[Caines et~al.(2022)Caines, Sisson, and Rowbottom]{caines20224dct}
Rhydian Caines, Naomi~K Sisson, and Carl~G Rowbottom.
\newblock 4dct and vmat for lung patients with irregular breathing.
\newblock \emph{Journal of Applied Clinical Medical Physics}, 23\penalty0 (1):\penalty0 e13453, 2022.

\bibitem[Chen et~al.(2022)Chen, Frey, He, Segars, Li, and Du]{chen2022transmorph}
Junyu Chen, Eric~C Frey, Yufan He, William~P Segars, Ye Li, and Yong Du.
\newblock Transmorph: Transformer for unsupervised medical image registration.
\newblock \emph{Medical image analysis}, 82:\penalty0 102615, 2022.

\bibitem[Chen et~al.(2024)Chen, Shi, Zheng, Yan, Hu, Zhu, and Mou]{chen2024ultrasound}
Tingxiu Chen, Yilei Shi, Zixuan Zheng, Bingcong Yan, Jingliang Hu, Xiao~Xiang Zhu, and Lichao Mou.
\newblock Ultrasound image-to-video synthesis via latent dynamic diffusion models.
\newblock In \emph{MICCAI}, pages 764--774. Springer, 2024.

\bibitem[Choi et~al.(2020)Choi, Kim, Han, Xu, and Lee]{choi2020channel}
Myungsub Choi, Heewon Kim, Bohyung Han, Ning Xu, and Kyoung~Mu Lee.
\newblock Channel attention is all you need for video frame interpolation.
\newblock In \emph{AAAI}, pages 10663--10671, 2020.

\bibitem[Choi et~al.(2021)Choi, Lee, Kim, and Lee]{choi2021motion}
Myungsub Choi, Suyoung Lee, Heewon Kim, and Kyoung~Mu Lee.
\newblock Motion-aware dynamic architecture for efficient frame interpolation.
\newblock In \emph{Proceedings of the IEEE/CVF International Conference on Computer Vision}, pages 13839--13848, 2021.

\bibitem[Danier et~al.(2024)Danier, Zhang, and Bull]{danier2024ldmvfi}
Duolikun Danier, Fan Zhang, and David Bull.
\newblock Ldmvfi: Video frame interpolation with latent diffusion models.
\newblock In \emph{Proceedings of the AAAI Conference on Artificial Intelligence}, pages 1472--1480, 2024.

\bibitem[Dosovitskiy et~al.(2015)Dosovitskiy, Fischer, Ilg, Hausser, Hazirbas, Golkov, Van Der~Smagt, Cremers, and Brox]{dosovitskiy2015flownet}
Alexey Dosovitskiy, Philipp Fischer, Eddy Ilg, Philip Hausser, Caner Hazirbas, Vladimir Golkov, Patrick Van Der~Smagt, Daniel Cremers, and Thomas Brox.
\newblock Flownet: Learning optical flow with convolutional networks.
\newblock In \emph{ICCV}, pages 2758--2766, 2015.

\bibitem[Ehrhardt et~al.(2013)Ehrhardt, Lorenz, et~al.]{ehrhardt20134d}
Jan Ehrhardt, Cristian Lorenz, et~al.
\newblock \emph{4D modeling and estimation of respiratory motion for radiation therapy}.
\newblock Springer, 2013.

\bibitem[Guo et~al.(2020)Guo, Bi, Ahn, Feng, Wang, and Kim]{guo2020spatiotemporal}
Yuyu Guo, Lei Bi, Euijoon Ahn, Dagan Feng, Qian Wang, and Jinman Kim.
\newblock A spatiotemporal volumetric interpolation network for 4d dynamic medical image.
\newblock In \emph{CVPR}, pages 4726--4735, 2020.

\bibitem[Heusel et~al.(2017)Heusel, Ramsauer, Unterthiner, Nessler, and Hochreiter]{heusel2017gans}
Martin Heusel, Hubert Ramsauer, Thomas Unterthiner, Bernhard Nessler, and Sepp Hochreiter.
\newblock Gans trained by a two time-scale update rule converge to a local nash equilibrium.
\newblock \emph{NeurIPS}, 30, 2017.

\bibitem[Ho et~al.(2020)Ho, Jain, and Abbeel]{ho2020denoising}
Jonathan Ho, Ajay Jain, and Pieter Abbeel.
\newblock Denoising diffusion probabilistic models.
\newblock \emph{Advances in neural information processing systems}, 33:\penalty0 6840--6851, 2020.

\bibitem[Ho et~al.(2022)Ho, Salimans, Gritsenko, Chan, Norouzi, and Fleet]{ho2022video}
Jonathan Ho, Tim Salimans, Alexey Gritsenko, William Chan, Mohammad Norouzi, and David~J Fleet.
\newblock Video diffusion models.
\newblock \emph{NeurIPS}, 35:\penalty0 8633--8646, 2022.

\bibitem[Hor et~al.(2011)Hor, Baumann, Pedrizzetti, Tonti, Gottliebson, Taylor, Benson, and Mazur]{hor2011magnetic}
Kan~N Hor, Rolf Baumann, Gianni Pedrizzetti, Gianni Tonti, William~M Gottliebson, Michael Taylor, D~Woodrow Benson, and Wojciech Mazur.
\newblock Magnetic resonance derived myocardial strain assessment using feature tracking.
\newblock \emph{Journal of visualized experiments: JoVE}, \penalty0 (48):\penalty0 2356, 2011.

\bibitem[Hore and Ziou(2010)]{hore2010image}
Alain Hore and Djemel Ziou.
\newblock Image quality metrics: Psnr vs. ssim.
\newblock In \emph{2010 20th international conference on pattern recognition}, pages 2366--2369. IEEE, 2010.

\bibitem[Hu et~al.(2022)Hu, Jiang, Liao, Xiao, Jiang, and Wang]{hu2022spatial}
Mengshun Hu, Kui Jiang, Liang Liao, Jing Xiao, Junjun Jiang, and Zheng Wang.
\newblock Spatial-temporal space hand-in-hand: Spatial-temporal video super-resolution via cycle-projected mutual learning.
\newblock In \emph{Proceedings of the IEEE/CVF conference on computer vision and pattern recognition}, pages 3574--3583, 2022.

\bibitem[Hu et~al.(2024)Hu, Jiang, Zhong, Wang, and Zheng]{hu2024iq}
Mengshun Hu, Kui Jiang, Zhihang Zhong, Zheng Wang, and Yinqiang Zheng.
\newblock Iq-vfi: Implicit quadratic motion estimation for video frame interpolation.
\newblock In \emph{Proceedings of the IEEE/CVF Conference on Computer Vision and Pattern Recognition}, pages 6410--6419, 2024.

\bibitem[Huang et~al.(2022)Huang, Zhang, Heng, Shi, and Zhou]{huang2022real}
Zhewei Huang, Tianyuan Zhang, Wen Heng, Boxin Shi, and Shuchang Zhou.
\newblock Real-time intermediate flow estimation for video frame interpolation.
\newblock In \emph{European Conference on Computer Vision}, pages 624--642. Springer, 2022.

\bibitem[Hugo et~al.(2016)Hugo, Weiss, Sleeman, Balik, Keall, Lu, and Williamson]{hugo2016data}
Geoffrey~D Hugo, Elisabeth Weiss, William~C Sleeman, Salim Balik, Paul~J Keall, Jun Lu, and Jeffrey~F Williamson.
\newblock Data from 4d lung imaging of nsclc patients.
\newblock \emph{(No Title)}, 2016.

\bibitem[Ilg et~al.(2017)Ilg, Mayer, Saikia, Keuper, Dosovitskiy, and Brox]{ilg2017flownet}
Eddy Ilg, Nikolaus Mayer, Tonmoy Saikia, Margret Keuper, Alexey Dosovitskiy, and Thomas Brox.
\newblock Flownet 2.0: Evolution of optical flow estimation with deep networks.
\newblock In \emph{CVPR}, pages 2462--2470, 2017.

\bibitem[Jain et~al.(2024)Jain, Watson, Tabellion, Poole, Kontkanen, et~al.]{jain2024video}
Siddhant Jain, Daniel Watson, Eric Tabellion, Ben Poole, Janne Kontkanen, et~al.
\newblock Video interpolation with diffusion models.
\newblock In \emph{CVPR}, pages 7341--7351, 2024.

\bibitem[Jeung et~al.(2012)Jeung, Germain, et~al.]{jeung2012myocardial}
Mi-Young Jeung, Philippe Germain, et~al.
\newblock Myocardial tagging with mr imaging: overview of normal and pathologic findings.
\newblock \emph{Radiographics}, 32\penalty0 (5):\penalty0 1381--1398, 2012.

\bibitem[Jiang et~al.(2018)Jiang, Sun, Jampani, Yang, Learned-Miller, and Kautz]{jiang2018super}
Huaizu Jiang, Deqing Sun, Varun Jampani, Ming-Hsuan Yang, Erik Learned-Miller, and Jan Kautz.
\newblock Super slomo: High quality estimation of multiple intermediate frames for video interpolation.
\newblock In \emph{Proceedings of the IEEE/CVF Conference on Computer Vision and Pattern Recognition}, pages 9000--9008, 2018.

\bibitem[Jin et~al.(2023)Jin, Wu, Chen, Chen, Koo, and Hahm]{jin2023unified}
Xin Jin, Longhai Wu, Jie Chen, Youxin Chen, Jayoon Koo, and Cheul-hee Hahm.
\newblock A unified pyramid recurrent network for video frame interpolation.
\newblock In \emph{CVPR}, pages 1578--1587, 2023.

\bibitem[Kalluri et~al.(2023)Kalluri, Pathak, Chandraker, and Tran]{kalluri2023flavr}
Tarun Kalluri, Deepak Pathak, Manmohan Chandraker, and Du Tran.
\newblock Flavr: Flow-agnostic video representations for fast frame interpolation.
\newblock In \emph{Proceedings of the IEEE/CVF winter conference on applications of computer vision}, pages 2071--2082, 2023.

\bibitem[Kim and Ye(2022)]{kim2022diffusion}
Boah Kim and Jong~Chul Ye.
\newblock Diffusion deformable model for 4d temporal medical image generation.
\newblock In \emph{MICCAI}, pages 539--548. Springer, 2022.

\bibitem[Kim et~al.(2024)Kim, Yoon, Park, Kim, and Yang]{kim2024data}
JungEun Kim, Hangyul Yoon, Geondo Park, Kyungsu Kim, and Eunho Yang.
\newblock Data-efficient unsupervised interpolation without any intermediate frame for 4d medical images.
\newblock In \emph{CVPR}, pages 11353--11364, 2024.

\bibitem[Kong et~al.(2022)Kong, Jiang, Luo, Chu, Huang, Tai, Wang, and Yang]{kong2022ifrnet}
Lingtong Kong, Boyuan Jiang, Donghao Luo, Wenqing Chu, Xiaoming Huang, Ying Tai, Chengjie Wang, and Jie Yang.
\newblock Ifrnet: Intermediate feature refine network for efficient frame interpolation.
\newblock In \emph{Proceedings of the IEEE/CVF Conference on Computer Vision and Pattern Recognition}, pages 1969--1978, 2022.

\bibitem[Li et~al.(2024)Li, Yang, Li, Lomax, Zhang, and Buhmann]{li2024cpt}
Xia Li, Runzhao Yang, Xiangtai Li, Antony Lomax, Ye Zhang, and Joachim Buhmann.
\newblock Cpt-interp: Continuous spatial and temporal motion modeling for 4d medical image interpolation.
\newblock \emph{arXiv preprint arXiv:2405.15385}, 2024.

\bibitem[Li et~al.(2023)Li, Zhu, Han, Hou, Guo, and Cheng]{li2023amt}
Zhen Li, Zuo-Liang Zhu, Ling-Hao Han, Qibin Hou, Chun-Le Guo, and Ming-Ming Cheng.
\newblock Amt: All-pairs multi-field transforms for efficient frame interpolation.
\newblock In \emph{CVPR}, pages 9801--9810, 2023.

\bibitem[Liang et~al.(2024)Liang, Yin, Xu, Liang, Wang, Plataniotis, Zhao, and Wei]{liang2024diffusion4d}
Hanwen Liang, Yuyang Yin, Dejia Xu, Hanxue Liang, Zhangyang Wang, Konstantinos~N Plataniotis, Yao Zhao, and Yunchao Wei.
\newblock Diffusion4d: Fast spatial-temporal consistent 4d generation via video diffusion models.
\newblock \emph{arXiv preprint arXiv:2405.16645}, 2024.

\bibitem[Liu et~al.(2024)Liu, Zhang, Li, Yan, Gao, Chen, Yuan, Huang, Sun, Gao, et~al.]{liu2024sora}
Yixin Liu, Kai Zhang, Yuan Li, Zhiling Yan, Chujie Gao, Ruoxi Chen, Zhengqing Yuan, Yue Huang, Hanchi Sun, Jianfeng Gao, et~al.
\newblock Sora: A review on background, technology, limitations, and opportunities of large vision models.
\newblock \emph{arXiv preprint arXiv:2402.17177}, 2024.

\bibitem[Lu et~al.(2022)Lu, Wu, Lin, Lu, and Jia]{lu2022video}
Liying Lu, Ruizheng Wu, Huaijia Lin, Jiangbo Lu, and Jiaya Jia.
\newblock Video frame interpolation with transformer.
\newblock In \emph{CVPR}, pages 3532--3542, 2022.

\bibitem[Meyer et~al.(2018)Meyer, Djelouah, McWilliams, Sorkine-Hornung, Gross, and Schroers]{meyer2018phasenet}
Simone Meyer, Abdelaziz Djelouah, Brian McWilliams, Alexander Sorkine-Hornung, Markus Gross, and Christopher Schroers.
\newblock Phasenet for video frame interpolation.
\newblock In \emph{Proceedings of the IEEE/CVF Conference on Computer Vision and Pattern Recognition}, pages 498--507, 2018.

\bibitem[Mizuno and Muto(2021)]{mizuno2021preoperative}
Kotaro Mizuno and Masahiro Muto.
\newblock Preoperative evaluation of pleural adhesion in patients with lung tumors using four-dimensional computed tomography performed during natural breathing.
\newblock \emph{Medicine}, 100\penalty0 (47):\penalty0 e27800, 2021.

\bibitem[Niklaus and Liu(2020)]{niklaus2020softmax}
Simon Niklaus and Feng Liu.
\newblock Softmax splatting for video frame interpolation.
\newblock In \emph{Proceedings of the IEEE/CVF conference on computer vision and pattern recognition}, pages 5437--5446, 2020.

\bibitem[Ouyang et~al.(2020)Ouyang, He, Ghorbani, Yuan, Ebinger, Langlotz, Heidenreich, Harrington, Liang, Ashley, et~al.]{ouyang2020video}
David Ouyang, Bryan He, Amirata Ghorbani, Neal Yuan, Joseph Ebinger, Curtis~P Langlotz, Paul~A Heidenreich, Robert~A Harrington, David~H Liang, Euan~A Ashley, et~al.
\newblock Video-based ai for beat-to-beat assessment of cardiac function.
\newblock \emph{Nature}, 580\penalty0 (7802):\penalty0 252--256, 2020.

\bibitem[Park et~al.(2023)Park, Kim, and Kim]{park2023biformer}
Junheum Park, Jintae Kim, and Chang-Su Kim.
\newblock Biformer: Learning bilateral motion estimation via bilateral transformer for 4k video frame interpolation.
\newblock In \emph{CVPR}, pages 1568--1577, 2023.

\bibitem[Roy et~al.(2023)Roy, Koehler, Ulrich, Baumgartner, Petersen, Isensee, Jaeger, and Maier-Hein]{roy2023mednext}
Saikat Roy, Gregor Koehler, Constantin Ulrich, Michael Baumgartner, Jens Petersen, Fabian Isensee, Paul~F Jaeger, and Klaus~H Maier-Hein.
\newblock Mednext: transformer-driven scaling of convnets for medical image segmentation.
\newblock In \emph{MICCAI}, pages 405--415. Springer, 2023.

\bibitem[Smiseth et~al.(2016)Smiseth, Torp, Opdahl, Haugaa, and Urheim]{smiseth2016myocardial}
Otto~A Smiseth, Hans Torp, Anders Opdahl, Kristina~H Haugaa, and Stig Urheim.
\newblock Myocardial strain imaging: how useful is it in clinical decision making?
\newblock \emph{European heart journal}, 37\penalty0 (15):\penalty0 1196--1207, 2016.

\bibitem[Sun et~al.(2018)Sun, Yang, Liu, and Kautz]{sun2018pwc}
Deqing Sun, Xiaodong Yang, Ming-Yu Liu, and Jan Kautz.
\newblock Pwc-net: Cnns for optical flow using pyramid, warping, and cost volume.
\newblock In \emph{CVPR}, pages 8934--8943, 2018.

\bibitem[Teed and Deng(2020)]{teed2020raft}
Zachary Teed and Jia Deng.
\newblock Raft: Recurrent all-pairs field transforms for optical flow.
\newblock In \emph{ECCV}, pages 402--419. Springer, 2020.

\bibitem[Tolstov(2012)]{tolstov2012fourier}
Georgi~P Tolstov.
\newblock \emph{Fourier series}.
\newblock Courier Corporation, 2012.

\bibitem[Unterthiner et~al.(2018)Unterthiner, Van~Steenkiste, Kurach, Marinier, Michalski, and Gelly]{unterthiner2018towards}
Thomas Unterthiner, Sjoerd Van~Steenkiste, Karol Kurach, Raphael Marinier, Marcin Michalski, and Sylvain Gelly.
\newblock Towards accurate generative models of video: A new metric \& challenges.
\newblock \emph{arXiv preprint arXiv:1812.01717}, 2018.

\bibitem[Wang et~al.(2009)Wang, Hayes, Paskalev, Jin, Buyyounouski, Ma, and Feigenberg]{wang2009dosimetric}
Lu Wang, Shelly Hayes, Kamen Paskalev, Lihui Jin, Mark~K Buyyounouski, Charlie C-M Ma, and Steve Feigenberg.
\newblock Dosimetric comparison of stereotactic body radiotherapy using 4d ct and multiphase ct images for treatment planning of lung cancer: evaluation of the impact on daily dose coverage.
\newblock \emph{Radiotherapy and Oncology}, 91\penalty0 (3):\penalty0 314--324, 2009.

\bibitem[Wang et~al.(2024)Wang, Lipson, and Deng]{wang2024sea}
Yihan Wang, Lahav Lipson, and Jia Deng.
\newblock Sea-raft: Simple, efficient, accurate raft for optical flow.
\newblock In \emph{ECCV}, pages 36--54. Springer, 2024.

\bibitem[Wei et~al.(2023)Wei, Kuo, Tseng, and Chen]{wei2023mpvf}
Tzu-Ti Wei, Chin Kuo, Yu-Chee Tseng, and Jen-Jee Chen.
\newblock Mpvf: 4d medical image inpainting by multi-pyramid voxel flows.
\newblock \emph{IEEE Journal of Biomedical and Health Informatics}, 2023.

\bibitem[You et~al.(2023)You, Ding, Zhang, Wu, Yu, Gu, and Yang]{you2023semantic}
Xin You, Ming Ding, Minghui Zhang, Yangqian Wu, Yi Yu, Yun Gu, and Jie Yang.
\newblock Semantic difference guidance for the uncertain boundary segmentation of ct left atrial appendage.
\newblock In \emph{MICCAI}, pages 121--131. Springer, 2023.

\bibitem[You et~al.(2024)You, He, Yang, and Gu]{you2024learning}
Xin You, Junjun He, Jie Yang, and Yun Gu.
\newblock Learning with explicit shape priors for medical image segmentation.
\newblock \emph{IEEE Transactions on Medical Imaging}, 2024.

\bibitem[You et~al.(2025{\natexlab{a}})You, Lou, Zhang, Yang, and Gu]{you2025slord}
Xin You, Yixin Lou, Minghui Zhang, Jie Yang, and Yun Gu.
\newblock Slord: Structural low-rank descriptors for shape consistency in vertebrae segmentation.
\newblock \emph{IEEE Journal of Biomedical and Health Informatics}, 2025{\natexlab{a}}.

\bibitem[You et~al.(2025{\natexlab{b}})You, Zhang, Zhang, Yang, and Navab]{you2025temporal}
Xin You, Minghui Zhang, Hanxiao Zhang, Jie Yang, and Nassir Navab.
\newblock Temporal differential fields for 4d motion modeling via image-to-video synthesis.
\newblock \emph{arXiv preprint arXiv:2505.17333}, 2025{\natexlab{b}}.

\bibitem[Zhang et~al.(2024)Zhang, Zheng, You, Zheng, and Gu]{zhang2024pass}
Chuyan Zhang, Hao Zheng, Xin You, Yefeng Zheng, and Yun Gu.
\newblock Pass: test-time prompting to adapt styles and semantic shapes in medical image segmentation.
\newblock \emph{IEEE Transactions on Medical Imaging}, 2024.

\bibitem[Zhang et~al.(2023)Zhang, Zhu, Wang, Chen, Wu, and Wang]{zhang2023extracting}
Guozhen Zhang, Yuhan Zhu, Haonan Wang, Youxin Chen, Gangshan Wu, and Limin Wang.
\newblock Extracting motion and appearance via inter-frame attention for efficient video frame interpolation.
\newblock In \emph{Proceedings of the IEEE/CVF Conference on Computer Vision and Pattern Recognition}, pages 5682--5692, 2023.

\bibitem[Zhang et~al.(2025)Zhang, Chen, Wang, Liu, Wang, and Qiao]{zhang20254diffusion}
Haiyu Zhang, Xinyuan Chen, Yaohui Wang, Xihui Liu, Yunhong Wang, and Yu Qiao.
\newblock 4diffusion: Multi-view video diffusion model for 4d generation.
\newblock \emph{NeurIPS}, 37:\penalty0 15272--15295, 2025.

\bibitem[Zhang et~al.(2018)Zhang, Isola, Efros, Shechtman, and Wang]{zhang2018unreasonable}
Richard Zhang, Phillip Isola, Alexei~A Efros, Eli Shechtman, and Oliver Wang.
\newblock The unreasonable effectiveness of deep features as a perceptual metric.
\newblock In \emph{CVPR}, pages 586--595, 2018.

\bibitem[Zhong et~al.(2024)Zhong, Krishnan, Sun, Qiao, Ma, and Wang]{zhong2024clearer}
Zhihang Zhong, Gurunandan Krishnan, Xiao Sun, Yu Qiao, Sizhuo Ma, and Jian Wang.
\newblock Clearer frames, anytime: Resolving velocity ambiguity in video frame interpolation.
\newblock In \emph{ECCV}, pages 346--363. Springer, 2024.

\bibitem[Zhou et~al.(2023)Zhou, Li, Han, and Lu]{zhou2023exploring}
Kun Zhou, Wenbo Li, Xiaoguang Han, and Jiangbo Lu.
\newblock Exploring motion ambiguity and alignment for high-quality video frame interpolation.
\newblock In \emph{CVPR}, pages 22169--22179, 2023.

\end{thebibliography}
}

\clearpage
\setcounter{page}{1}
\maketitlesupplementary

\section{Dataset Settings}
\label{sec:rationale}
\noindent \textbf{ACDC.}  The ACDC dataset contains 80\% pathological cardiac cases, including pathologies with myocardial infarction, cardiomyopathy. All MRI volumes are resampled with a voxel space of $1.5 \times 1.5 \times 3.12mm^{3}$. Besides, all cardiac  scans have been cropped with a centered patch. The patch size is set as $128 \times 128 \times 32$. The frame number $N$ shows a range of $[6, 16]$. Min-max scaling at $[0, 1]$ is applied to ensure consistent scaling across all scans.
\\

\noindent \textbf{4D Lung.} In the case of the 4D-lung dataset, the models are trained to predict the four intermediate frames ($10\%,
20\%, 30\%, 40\%$) between the end-inspiratory $(0\%)$ and
end-expiratory $(50\%)$ phases. Only CT images captured using kilovoltage energy are included in the study due to
their superior image quality. The data preprocessing strategy is the same as that in \cite{kim2024data}.

\section{Implementation Details}

\noindent \textbf{Network Details.} For the first stage, the VAE is not to regulate the whole pipeline, but to utilize a MedNeXt \cite{roy2023mednext} structure for encoding temporal features and learning Fourier bases. The VAE maps the image space into the downsampled latent space with a ratio of $1/8$. Specifically, the core component for MedNeXt is the MedNeXtBlock. For more details of the VAE, please refer to the source codes released \href{https://anonymous.4open.science/r/FB-Diff-DC98}{here}. For the latent diffusion UNet, we select a more lightweight MedNeXt as the baseline, with the downsampling scale equal to $1/4$. The diffusion timestep is set as $1000$. $L_{2}$ norm is chosen as the loss function for the diffusion process.
\\

\noindent \textbf{Training Details.}
All models are trained using AdamW optimizer with the linear warm-up strategy. For the taining of VAE, the initial learning rate is set as $3e\mbox{-}4$ with a cosine learning rate decay scheduler, and weight decay is set as $1e\mbox{-}5$. While for the training of the diffusion model, the learning rate is set as $1e\mbox{-}4$. The batch size is set as $2$. Experiments are implemented based on Pytorch and 2 NVIDIA RTX 4090 GPUs.

\begin{table}[!t]
  \begin{center}
    \caption{\footnotesize{(a) Generalization on cardiac ultrasound in EchoNet-Dynamic \cite{ouyang2020video}. (b) Model efficiency on training time, computational costs, and per-case inference speed.}}
\label{model_efficiency}
    \centering
    \resizebox{1.1\columnwidth}{!}{
    \begin{tabular}{lcccccc}
        \hline
    \multirow{2}*{Model} & \multicolumn{3}{c}{(a) Cardiac ultrasound} & \multicolumn{3}{c}{(b) Model efficiency} \\
      \cmidrule(r){2-4}    \cmidrule(r){5-7}
    & PSNR (dB) $\uparrow$ & LPIPS $\downarrow$ &  FVD $\downarrow$ &  Training time (h) & FLOPs (T) & Inference (s) \\
            \hline
        Voxelmorph [2] & 28.40 & 2.492 & 295.3 & 5.6 & 0.49 & 1.09 \\  
        IFRNet [31] & 29.95 & 2.017 & 261.8 & 21.3 & 1.92 & 1.27 \\ 
        UVI-Net [30] & \textbf{30.87} & \underline{1.818} & \underline{243.7} & 18.5 & 1.27 & 0.63 \\
        Conditional diff [16] & 26.67 & 2.578 & 337.2 & 12.4 & 2.37 & 37.80 \\ 
        FB-Diff & \underline{30.51} & \textbf{1.654} & \textbf{227.0} & 18.0 & 1.58 & 29.50 \\
        \hline 
    \end{tabular}}
  \end{center}
\end{table} 

\section{Model Efficiency}
We have added the model efficiency metrics. Table \ref{model_efficiency} reports the training time, FLOPs, and per-case inference speed for models. Overall, FB-Diff offers a good trade-off in performance and model efficiency.

\section{Generalization to other modalities}
We tested FB-Diff on a different imaging modality to confirm generality. Using the cardiac ultrasound dataset proposed by \cite{ouyang2020video}, FB-Diff achieves comparable or better performance than benchmark methods. As revealed in Table \ref{model_efficiency}, FB-Diff achieves better temporal consistency for interpolated videos while maintaining promising reconstruction metrics.

\clearpage
\setcounter{page}{2}

\begin{figure*}[!t]
\centerline{\includegraphics[width=0.96\linewidth]{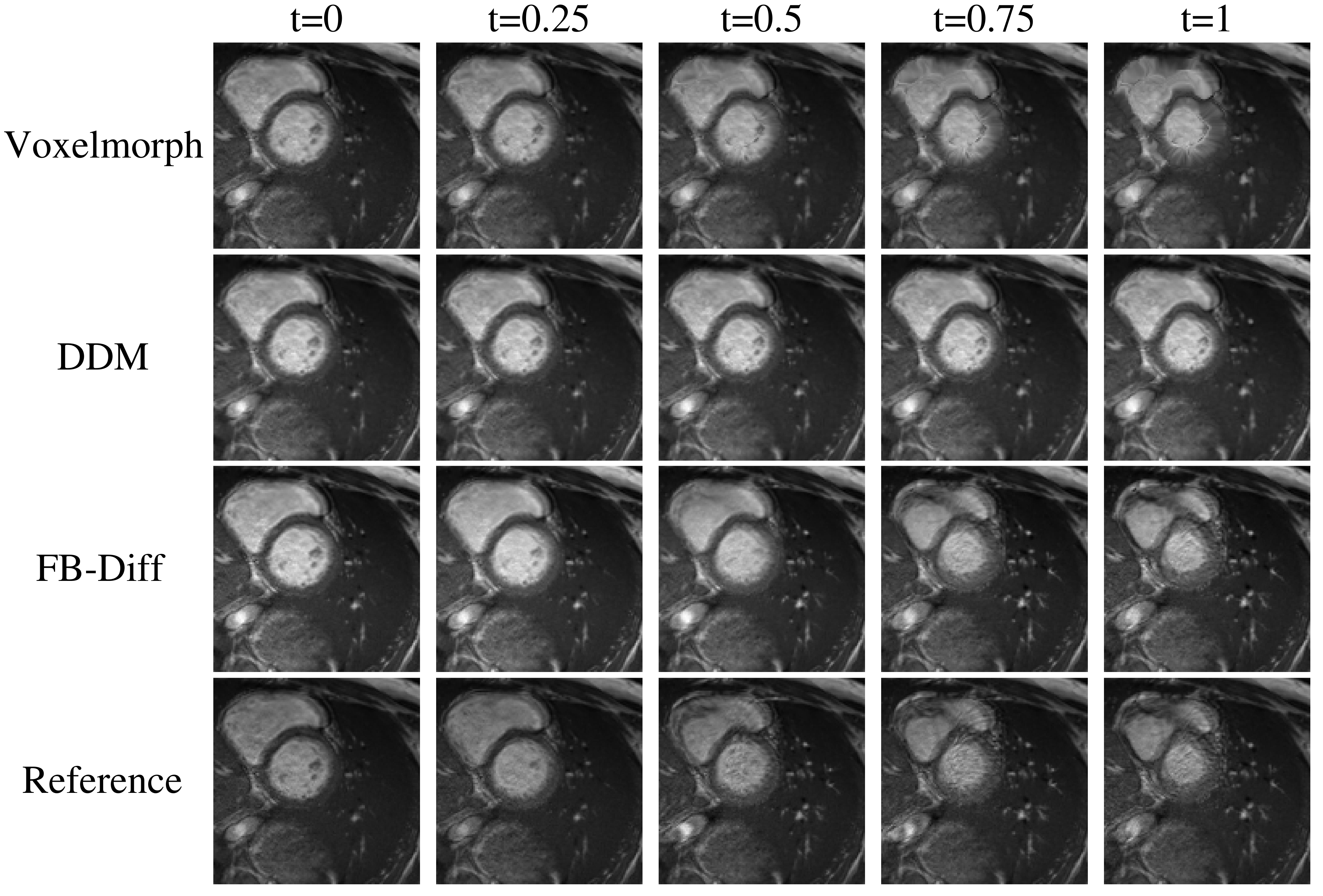}}
\caption{Temporal variation comparison between FB-Diff and existing methods with the linear motion hypothesis.}
\label{temporal_supple}
\end{figure*}

\begin{figure*}[!t]
\centerline{\includegraphics[width=0.96\linewidth]{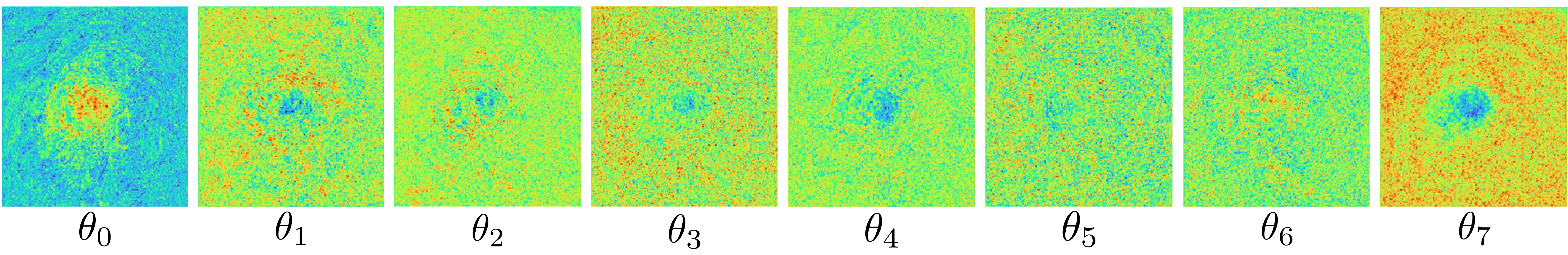}}
\caption{The spectral intensity visualizations of the first eight well-learned physiology motion priors on ACDC.}
\label{physiology motion}
\end{figure*}

\end{document}